\documentstyle[12pt]{article}
\begin{document}
\def\etaten{\eta_{10}}
\def\he{${\rm {}^4He\:\:}$}
\def\be{\begin{equation}}
\def\ee{\end{equation}}
\def\bc{\begin{center}}
\def\ec{\end{center}}
\def\beq{\begin{eqnarray}}
\def\eeq{\end{eqnarray}}
\def\beqd{\begin{eqnarray*}}
\def\eeqd{\end{eqnarray*}}
\def\nin{\noindent}
\def\lra{$\leftrightarrow$ }
\def\eset{$\not\!\!\:0$ }
\def\bull{$\bullet$}
\def\reaceight{$^3$He$(\alpha , \gamma )^7$Be}
\def\reacnine{$^3$H$(\alpha , \gamma )^7$Li}

\pagestyle{empty}
\rightline{{\bf CWRU-P2-94}}
\rightline{January 1994}
\baselineskip=16pt
\vskip 0.5in
\begin{center}
\bf\large {REFINED BIG BANG NUCLEOSYNTHESIS CONSTRAINTS ON  
$\Omega_{Baryon}$
 and
$N_{\nu}$} 

\end{center}
\vskip0.2in
\begin{center}
Peter J. Kernan
and
Lawrence M. Krauss\footnote{Also Dept of Astronomy}
\vskip .1in
 {\small\it Department of Physics\\
Case Western Reserve University\\ 10900 Euclid Ave., Cleveland, OH
44106-7079}
\vskip 0.4in
\end{center}
\centerline{{\bf Abstract}}
\noindent
We include correlations between elemental
abundances in a Monte Carlo statistical analysis of BBN predictions,  
which,
along with updated reaction rates and an improved BBN code, lead to  
tightened
constraints on $\Omega_{B}$ and $N_{\nu}$. Observational upper limits
on the primordial \he and $\rm {D + {^3He}}$ fractions of $24 \%$ (by  
weight)
and $ 10^{-4}$ respectively lead to the limits:
 $0.015 \le \Omega_{B} \le 0.070$, and $N_{\nu} \le 3.04$. The
former limit appears to be incompatible with purely baryonic galactic  
halo
dark matter, while the latter puts qualitatively new constraints on
neutrinos, and physics beyond the standard model.

\newpage
\pagestyle{plain}
\setcounter{page}{1}
\baselineskip=21pt
  The
remarkable agreement, both qualitative and quantitative, of the  
predicted
primordial light element abundances and those inferred from present
observations yields some of the strongest evidence in favor of a
homogeneous FRW Big Bang cosmology.  Because of this, significant  
efforts have
taken place over 20 years to refine BBN predictions, and the
observational constraints they are related to.  Several factors have
contributed to the maturing of this field, including the
incorporation of elements beyond \he in comparison between theory and
observation (i.e. \cite{who}), and more recently: an
updated BBN code \cite{kawano}, a more accurate measured neutron half
life\cite{helium}, new estimates of the actual primordial \he, $D
+^3He$, and $^7Li$ abundances \cite{walker,con}, and finally the
determination of BBN uncertainties via Monte Carlo analysis
\cite{kraussrom}.  All of these, when combined together  
\cite{smithetal}
yield a consistent picture of homogeneous BBN which is at the same  
time
strongly constrained by observation.

We have returned to re-analyze BBN constraints
motivated by three factors:  new measurements of 

several BBN reactions, the development of an improved BBN code,
and finally the realization that a correct statistical determination  
of BBN
predictions should include correlations between the different  
elemental
abundances. Each serves to further restrict the
allowed range of the relevant cosmological observables  
$\Omega_{Baryon}$
and $N_{\nu}$.

\vskip 0.2 in

\noindent { \bf 1. New BBN Reaction Rates:}  By far the most  
accurately
measured BBN input parameter is the neutron half-life, which governs
the strength of the weak interaction which interconverts neutrons and
protons, and which in turn helps govern when this
reaction drops out of equilibrium.  Since this effectively
determines the abundance of free neutrons at the onset of BBN, it is  
crucial
in determining the remnant abundance of \he.  With the advent of  
neutron
trapping, the uncertainty in the neutron half-life quickly
dropped to less than $0.5 \%$ by 1990.  Nevertheless, it is the  
uncertainty
in this parameter that governs the uncertainty in the predicted \he
abundance.  The world average
for the neutron half-life is now $\tau_N = 889
\pm 2.1 sec$ \cite{helium}, which has an uncertainty which is almost  
twice as
small as that used in previous published BBN analyses
\cite{walker,kraussrom,smithetal}.  We utilize the updated value in  
our
analysis.

 Next, a recent measurement of 

 $^7Be + p \rightarrow  \gamma + ^8B $
suggests\cite{gai} a rate about $20 \%$ smaller at low energies than  
previous
estimates.  One might expect that at high values of 

$\etaten$ (defined by the relation $\Omega_{B} =.0036 h^{-2}  
(T/2.726)^3
\etaten \times 10^{10}$, where $T$ is the microwave background  
temperature
today, and $h$ defines the Hubble parameter by $ H= 100h$ km/(Mpc  
sec))
lowering this rate would result in less $^7Be$ destruction, which  
would
increase the $^7Li$ abundance resulting from the decay of $^7Be$
after BBN.  However, this is a subdominant destruction process for  
$^7Be$. 

We find that the reducing the rate by $20 \%$ in our code alters the  
remnant
$^7Li$ abundance by less than one part in $10^5$!

Outside of these, we updated the Kawano code to use the
reaction rates and uncertainties from in Smith { \it et al}
\cite{smithetal}. 

\vskip 0.2in

\noindent {\bf 2. New BBN Monte Carlo:}  Because
of the new importance of small corrections to the \he abundance when
comparing BBN predictions and observations, increased attention has  
been paid
recently to effects which may alter this abundance at the $1 \%$  
level or
less.  In our BBN Code several such effects were incorporated,  
resulting in
an $\etaten$ -independent correction of $+.0006$ to the lowest order  
value of
$Y_p$ (the \he mass fraction).  This is a change of $+.0031$ compared  
to the
value used in previous published
analyses\cite{walker,kraussrom,smithetal}.

This earlier value was based on correcting the lowest order value of
$Y_p$ by an amount $-.0025$ \cite{kawano}, based on the work of Dicus  
{\it
et al} \cite{dic:accbbn}.  The Dicus {\it et
al} correction has two significant pieces: -.0013 from integrating  
the weak
rates rather than using an expansion in powers of $T$ to calculate
$\lambda(n\leftrightarrow p)$, and 

-.0009 from using the correct Coulomb correction, rather than simply
scaling the neutron lifetime. This latter approximation 

incorrectly ``Coulomb corrects'' rates which do not feel the  
electromagnetic
potential, such as $ne^+\to p\bar{\nu}$, and also ignores any  
temperature
dependence. The
remaining corrections --- radiative, finite
temperature, electron mass effects and neutrino heating---either 

effectively cancel
(the first two) or are insignificant (the last two)  
\cite{dic:accbbn}.

In the present code, more than half of the new correction  

is due to finer integration of the nuclear
abundances. Making the time-step in the
code short enough that different Runge-Kutta drivers result in the  
same
number for the \he abundance produces a nearly $\etaten$ independent
change in $Y_p$ of +.0017 \cite{kernan}. The other major change is  
the
inclusion of ${M_N^{-1}}$ effects\cite{seck:accbbn}. Seckel
showed that the effects on the weak rates due to nucleon recoil, weak
magnetism, thermal motion of the nucleon target and time dilation of
the neutron lifetime combine to increase $Y_p$ by $\sim$ .0012. 

The kinematics of nuclear recoil, are responsible for roughly 

25\% of the increase which Seckel found \cite{kernan}.
Also included in the correction
is an small increase of $.0002$ in $Y_p$ from momentum dependent  
neutrino 

decoupling \cite{turndod,ftd:accbbn}.

Finally, we have utilized a Monte Carlo procedure in order to  
incorporate
existing uncertainties and determine confidence limits on parameters.   
Such a
procedure was first carried out by Krauss and Romanelli  
\cite{kraussrom}, who
chose the BBN reaction rates from a (temperature-independent)  
distribution
based on then existing experimental uncertainties. Their procedure  
was
further refined by Smith et al \cite{smithetal}, who both updated the
experimental uncertainties, and utilized temperature dependent
uncertainties in their analysis.  Here we utilized the nuclear
reaction rate uncertainties quoted by Smith et al (including the
temperature dependent uncertainties for
 \reaceight $\:$ and \reacnine) except for the updated reactions
described earlier. Each reaction rate was determined using a Gaussian
distributed random variable
centered on unity, with a width based on the
quoted  $1-\sigma$ uncertainty from Smith {\it et al}. For the 

rates without temperature
dependent uncertainties this number was used as a multiplier 

throughout the
nuclear abundance integration. For the two
rates with temperature dependent uncertainties the original uniformly
distributed random number was saved and mapped into a new gaussian 

distribution with
the appropriate width
at each time step. 

While the Smith {\it et al} analysis cut off each distribution at  
$\pm
2.6\sigma$, we did not made such a restriction.  Our code was  
designed to
generate warnings, discard data, and reset random numbers if reaction  
rate
values which were generated became nonsensical (i.e. negative).  
Warnings
were generated only for the temperature dependent rates. For  
\reacnine \  1
warning per 4000 BBN runs was generated, less than 1 warning per  
30000 runs
was generated for \reaceight.

The results of our updated BBN Monte Carlo analysis are displayed in  
figure
1, where the symmetric $95 \%$ confidence level predictions for each
elemental abundance are plotted. Also shown are claimed upper limits
for each of the light elements \cite{walker,con,smithetal} based on 

observation{\footnote{Where the estimates differ, we have used the  
more
conservative one.}}.   This figure also allows one to assess the
significance of the size of the corrections we have used in relation  
to the
width of the 95\% C.L. band for
$Y_p$, which turns out to be
$\sim$ .002. The total change in $Y_p$ of $\approx +.003$ from  
previous BBN
analyses conspires with the reduced uncertainty in
the neutron lifetime, which narrows the uncertainty in
$Y_p$ and also feeds into  the uncertainties in the other light  
elements,
 to reduce  the acceptable range where the
predicted BBN abundances are consistent with the inferred
primordial abundances.
\vskip 0.1 in

\noindent {\bf 3. Statistical Correlations Between Predicted  
Abundances}:
While the introduction of a Monte Carlo procedure was a significant
improvement in the evaluation of BBN
uncertainties and predictions, the determination of
limits on the allowed range of BBN parameters $\Omega_{baryon}$ and
$N_{\nu}$ based on comparison of symmetric $95 \%$ confidence limits
for single elemental abundances with observations, as has become the
standard procedure, overestimates the allowed range.  This is because  
the
BBN reaction network ties together all reactions, so that the  
predicted
elemental abundances are not statistically independent. In
addition, the use of symmetric confidence limits is too conservative.
Addressing both of these factors is a central feature of our work.

Figure 2 displays the locus of predicted values for the fractions  
$Y_p$
and D +$^3$He/H for 1000 BBN models generated from the distributions
described above for $\etaten =2.71$ (figure a) and $\etaten =3.08$  
(figure
b).  Also shown is the $\chi^2 =4$ joint confidence level contour  
derived
from this distribution, in a Gaussian approximation, calculating  
variances
and covariances in the standard manner.  The horizontal and vertical
tangents to this contour correspond to the individual symmetric $\pm
2\sigma$ limits on Gaussianly distributed $ x$
and $y$ variables. As can be seen, the distribution is close to  
Gaussian,
but has deviations.  Nevertheless, this approximation is useful to
quantify the magnitude of correlations and variances. We
calculated the normalized covariance matrices at different values of
$\etaten$, and display the covariances in Table 1.  As is evident  
from this
table, as well as the figure, and as is also well known on the basis  
of
analytical arguments, there is a strong anti-correlation between  
$Y_p$ and the
remnant D +$^3$He abundance.   Thus, those models where \he
is lower than the mean, and which therefore may be allowed by the
existing quoted upper bound of $24 \%$ on $Y_p$, will also generally  
produce
a larger remnant D+$^3$He/H abundance, which can be in conflict with  
the
quoted upper bound on this combination of $10^{-4}$. This will have  
the
effect of reducing the parameter space which is consistent with both  
limits,
as we now describe.

Because our Monte Carlo
generates the actual distribution of abundances, Gaussian or not, we
 determine
a $95 \% $ confidence limit on the allowed range of  
$\etaten(N_{\nu})$
by requiring that at least 50 models
out of 1000 lie within the joint range bounded by both the \he and D  
+$^3$He
upper limits, as shown in figure 2. This
is to be compared with the procedure which one would follow without
considering joint probability distributions.  In this case, one would
simply check whether 50 models lie {\it either} to the right of the 

D +$^3$He constraint for low $ \etaten$ (figure a), or below the \he
constraint for high $\etaten$ (figure b).  This is clearly a looser
constraint than that obtained using the joint distribution.  Finally,
the procedure which has been used to-date, which is to check whether  
the
symmetric $2 \sigma$ confidence limit for a single elemental  
abundance
crosses into the allowed region gives even a looser constraint, as  
can be
seen in figure 2a. This, after all corresponds to checking whether  
only 25
models lie to the right of the  $D +^3He$ constraint for low $  
\etaten$
(figure a), or below the \he constraint for high $\etaten$ (figure  
b).

In table 2 and figure 3 we display our results.  Here we show the $95  
\%$
confidence limits on $\etaten$, as we have defined them above, and  
also using
the looser procedures which ignore correlations.  As can be seen,
accounting for the correlations in the non-symmetric $95 \%$  
confidence limit
tightens constraints.  Moreover, the impact of the procedure becomes
stronger as the number of effective light neutrino species, $N_{\nu}$  
is
increased. Greater than $3.04$ effective light neutrino types is  
ruled out
only once correlations are taken into account.

We also determined an
upper limit on
$\etaten$ using just $^7Li$. Requiring
$^7Li/H
\le 2.3 \times 10^{-10}$ \cite{con,smithetal} yields a limit $\etaten
\le 5.27$ .  This is weaker than the \he limit, and there remains
some debate about the actual observational upper limit on
primordial $^7Li$ (i.e. changing 2.3 to 1.4\cite{walker} will lower  
the limit
on $\etaten$ to 4.15.).  Alternatively, we can use the bound on  
$\etaten$
derived above to set an allowed range of $9 \times  
10^{-11}\rightarrow 1.5
\times 10^{-10}$ on the primordial values of
$^7Li$, which should be compared with the
observational estimates.

\vskip 0.1 in

\noindent {\bf 4. Conclusions and Implications:}  The new 

constraints we have derived here on $\etaten$, and $N_{\nu}$, taken  
at
face value, have significant implications for cosmology, 

dark matter, and particle physics.  The limit on $\etaten$  
corresponds to
the limit $0.015 \le \Omega_{Baryon} \le 0.070$.  (To derive this  
bound we
required $ 0.4 \le h \le 0.8$, as is required by direct measurements  
and 

limits on the age of the universe.) Thus, { \it
if the quoted observational upper limits on the $Y_p$ and D+$^3$He/H
are valid}  homogeneous BBN implies that:
\vskip 0.05 in

\noindent (a)  The upper limit on $\Omega_{baryon}$ seems 

incompatible with all, or even galactic halo dark matter being
purely baryonic.  

\vskip 0.05 in

\noindent (b)  The bound on the number of effective light degrees of
freedom during nucleosynthesis is {\it very severe}, corresponding to  
less
than $0.04$ extra light neutrinos.  This is a qualitatively different
constraint than the previously quoted limit of $0.3$ extra neutrinos.   
For
example, it rules out {\it any} Dirac mass for a neutrino without  
some
extension of the standard model because even a right
handed component which freezes out at temperatures in
excess of $ 300$ GeV will contribute in excess of $0.047$ extra
neutrinos during BBN i.e. \cite{Kolbturn} without extra particles 

introduced whose annihilation can further suppress its abundance.   
Even
allowing  $0.047$ extra light neutrinos,
the upper limit on a Dirac mass would be reduced to $ \approx 5$  keV
\cite{fullermal,krauss}. Similarly, new light scalars are ruled out  
unless
they decouple above $300 GeV$.
 A $\nu_{\tau}$ mass greater than 0.5 MeV with lifetime exceeding 1  
sec.
is also ruled out due to its effect on the
expansion rate during BBN \i.e. see \cite{turner,kkkssw}.  Also,
neutrino interactions
 induced by  extended technicolor at scales less than $O(100)$ TeV  
are
ruled out \cite{kraternap}.  Moreover, sterile
right handed neutrinos \cite{dodelmal} would be ruled out as warm  
dark matter
as the lower limit on their mass would now be
$O(1 keV)$.   ( We will explore these constraints further in  
subsequent
work). 

\vskip 0.1in

Finally, having devoted considerable effort to accounting for
the statistical uncertainties in BBN predictions, we must still  
stress that
the largest, and most significant, uncertainties in the comparison of  
BBN
predictions with observations come from the latter.  Moreover, the
uncertainties in these observational limits are dominated by
systematic, and not statistical effects.  Hence, the $ 95 \%$  
confidence
limits we derive must be qualified by the recognition that their  
significance
is really only as good as the observational limits are. Such limits
cannot, at present, be taken to imply  statistically inviolate  
constraints on
neutrino parameters or $\Omega_B$.  In other words, systematic errors  
in the
quoted upper limits on the inferred primordial light element  
abundances could
allow the limits quoted here to be broadened.

Nevertheless, the theory can be carefully tested.  If, for
example, baryonic dark matter is found to make up the galactic halo  
or if
neutrino mass measurements conflict with the above bounds, this would
most likely imply that the quoted upper limits on \he, or on $^3He +  
D$ are
flawed.  This would be of great interest for stellar evolution  
studies. 

Indeed, the existing constraints from BBN are now so tight---  
requiring a
primordial \he fraction in excess of $23.8 \%$ for
consistency---that an agnostic view is prudent
at the present time as to whether the constraints derived
above will be satisfied or else whether observations will require  
revision in
the inferred primordial abundance estimates.  Finally, we note that
inhomogeneous BBN is not likely to alter this conclusion, as recent
work has established. \cite{fuller}.

\newpage

\begin{center}
{Table 1: Normalized Covariances} 

\vskip .2truein
\begin{tabular}{|l||c|c|c|c|c|c|}  \hline 

{}&\multicolumn{6}{|c|}{$\eta_{10}$} \\ \cline{2-7}
{covariance of}& 1.00 & 2.00 & 2.50 & 3.00 & 3.40 & 4.00 \\ \hline
$^4$He vs D + $^3$He & -.71 & -.60 & -.49 & -.47 & -.44 & -.6 \\
$^4$He vs $^7$Li & -.23 & .04 & .12 & .25 & .30 & .25 \\ \hline 

\end{tabular}
\end{center}

\vskip 0.3 in

\bc
{Table 2: Correlations \& $\eta_{10}$ limits}
\vskip .25 truein
\begin{tabular}{|l||c|c|c|c|}  \hline 

{95\% C.L.}&\multicolumn{4}{|c|}{$N_{\nu}$} \\ \cline{2-5}
 $\eta_{10}$ range & 3.0 & 3.025 & 3.04 & 3.05 \\ \hline
w/ corr. & 2.69 \lra 3.12 & 2.75 \lra 2.98 & 2.83 \lra 2.89 & 

\eset \\ \cline{1-1}
w/out corr. & 2.65 \lra 3.14 & 2.65 \lra 3.04 & 2.69 \lra 2.99  & 

2.69 \lra 2.95 \\ \cline{1-1}
sym. w/out corr. & 2.62 \lra 3.17 & 2.63 \lra 3.10 & 2.65 \lra 3.03 &
 2.66 \lra 3.00 \\ \hline
\end{tabular}
\ec

\newpage

\clearpage
\noindent {\bf Figure Captions}

\vskip 0.2in

\noindent Figure 1: BBN Monte Carlo predictions as a function of
$\etaten$.  Shown are symmetric  $95 \%$ confidence limits on each  
elemental
abundance.  Also shown are claimed upper limits inferred from  
observation.  

\vskip 0.1 in

\noindent Figure 2: Monte Carlo BBN predictions for $Y_p$ vs $D +  
^3He$ and
allowed range for (a) $\eta_{10} =2.71$, and (b) $\eta_{10} =3.08$. 

In (a) a Gaussian contour with $\pm 2 \sigma$ limits on each  
individual
variable is also shown.
\vskip 0.1 in
\noindent Figure 3: Number of models (out of 1000 total models) 

which satisfy constraints
$Y_p \le 24 \%$ and $ D + ^3He/H \le 10^{-4}$ as a function of $  
\etaten$, for
$3.0, 3.025, 3.04, 3.05$ effective light neutrino species.  Curves  
are
smoothed splines fit to the data. 

\clearpage

\vskip .2in

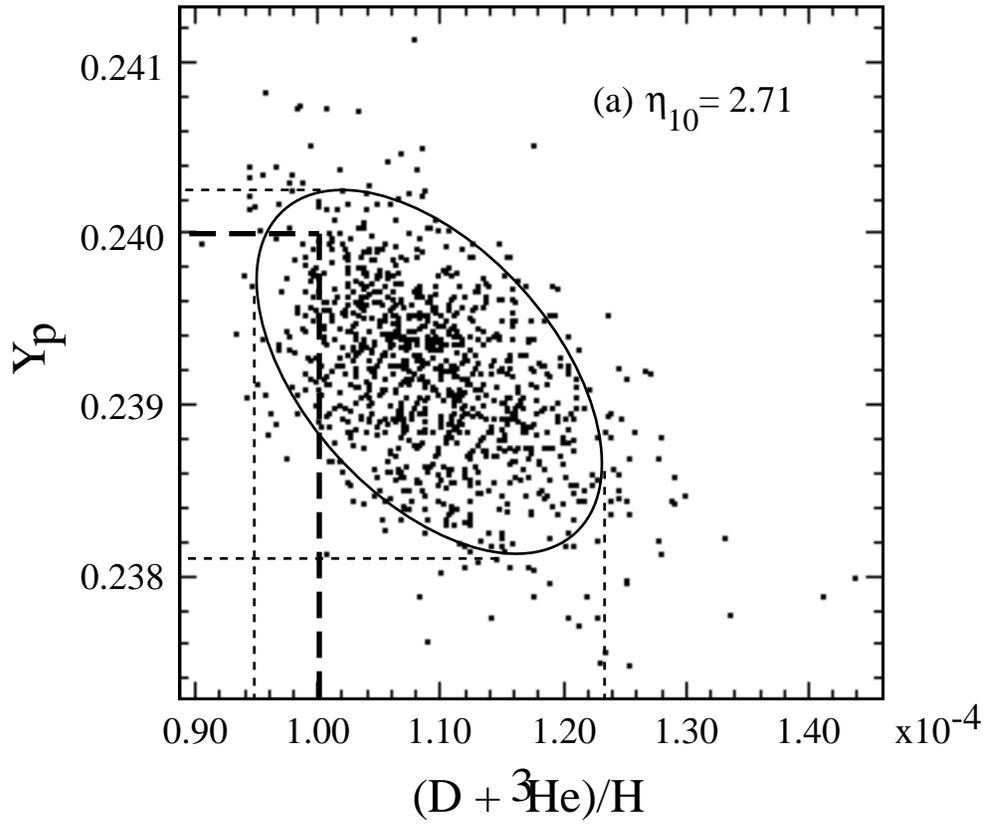
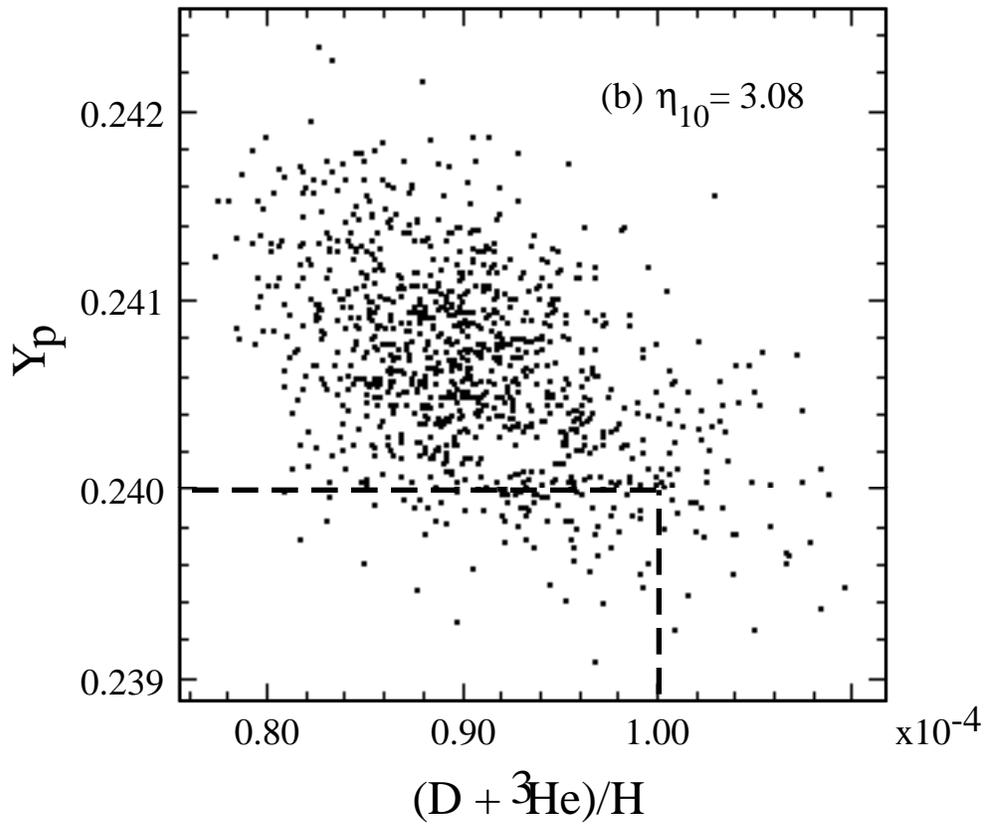
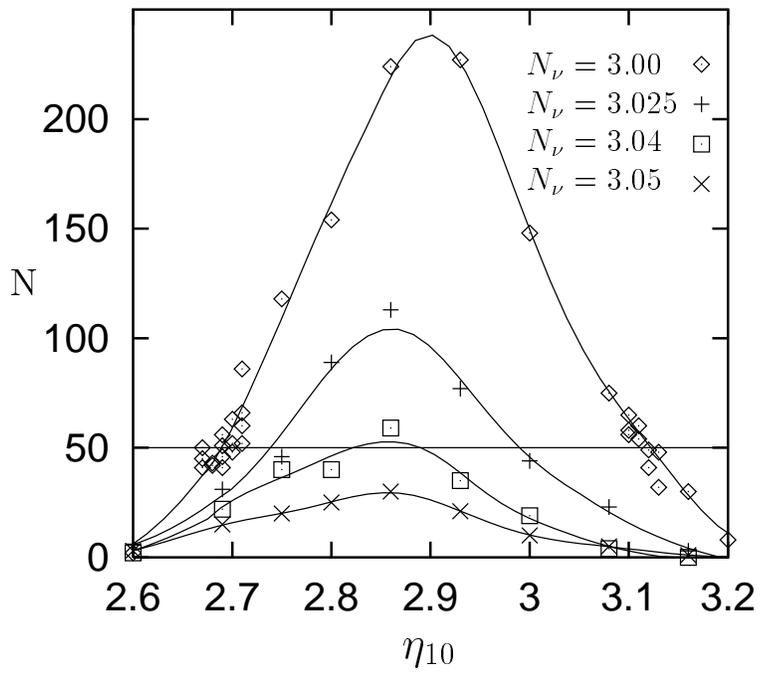
\end{document}